\documentclass{vgtc}                          

\ifpdf
  \pdfoutput=1\relax                   
  \usepackage{graphicx}                
  \DeclareGraphicsExtensions{.pdf,.png,.jpg,.jpeg} 
\else
  \usepackage{graphicx}                
  \DeclareGraphicsExtensions{.eps}     
\fi%

\graphicspath{{figures/}{pictures/}{images/}{./}} 
\usepackage{enumitem}
\usepackage{microtype}                 
\PassOptionsToPackage{warn}{textcomp}  
\usepackage{textcomp}                  
\usepackage{mathptmx}                  
\usepackage{times}                     
\usepackage{cite}                      
\usepackage{tabu}                      
\usepackage{booktabs}                  

\onlineid{0}

\vgtccategory{Research}

\vgtcinsertpkg

\title{Understanding and Measuring the Effects of \\Graphical Dimensions on Viewers' Perceived Chart Credibility}

\author{Hayeong Song\thanks{e-mail: hsong300@gatech.edu}\\ %
        \scriptsize Georgia Institute of Technology %
\and John Stasko\thanks{e-mail: stasko@cc.gatech.edu}\\ %
        \scriptsize Georgia Institute of Technology} %

\abstract{Journalists and visualization designers include visualizations in their articles and storytelling tools to deliver their message effectively. But design decisions they make to represent information, such as the graphical dimensions they choose and the viewer's familiarity with the content can impact the viewer's perceived credibility of charts. Especially in a context where little is known about sources of online information. But there is little experimental evidence that designers can refer to make decisions. 
Hence, this work aims to study and measure the effects of graphical dimensions and people's familiarity with the content on viewers' perceived chart credibility. I plan to conduct a crowd-sourced study with three graphical dimensions conditions, which are traditional charts, text annotation, and infographics. Then I will test these conditions on two user groups, which are domain experts and non-experts.  With these results, this work aims to provide chart guidelines for visual designers with experimental evidence. } 

\CCScatlist{
  \CCScatTwelve{Human-centered computing}{Visu\-al\-iza\-tion}{Empirical studies in visualization}{}{};
  \CCScatTwelve{Human-centered computing}{Visu\-al\-iza\-tion}{Visualization application domains}{Information visualization}{}
}

\begin{document}


\firstsection{Introduction}

\maketitle

Visualization designers, journalists, scholars, and system designers consider different visual representations to deliver a message or stories they produce, such as in storytelling tools. When designing, visual designers have to make a decision on what information (e.g., scientific evidence) to include and make design decisions to represent that information (e.g., infographics). These design decisions can impact viewers' perceived chart credibility. Prior works have shown that depending how information is visually represented can impact the viewer's perceived data quality~\cite{song2018s, song2021understanding} and data interpretation. For example, when the title and visualization were misaligned, viewers trusted the chart less which even lead to biases in their data interpretation~\cite{kong2018frames,kong2019trust}. Prior work has also shown that data is personal and people's background (e.g., where they are from, experiences, expertise) can impact people's perception and interpretation of charts~\cite{peck2019data}. However, there is little experimental evidence to guide designers as to what graphical dimensions and user characteristics (e.g., domain experts and novices) impact viewers' perceived chart credibility. Thus, in my work, I aim to understand and measure the effects of graphical dimensions on viewer's perceived chart credibility. Also, I aim to measure the effects of subject matter expertise  on their perceived credibility on charts. Based on these findings I will compose chart design guidelines that visual designers can refer to with experimental evidence for creating visualizations.

\section{Background and Related work}

Considering today's media environment online information is increasingly prevalent including newspapers, web blogs, and social media. Often time visualizations are used to represent this online information in storytelling tools or journalism to provide data-driven messages~\cite{knight2015data,loosen2020data,young2018makes}. But online information is not easily identified, which makes boundaries between perceived sources vague and ambiguous~\cite{appelman2016measuring,metzger2013credibility}. For example, online users view an article based on the search results that a search engine returns. But they view these online articles without contextual information and the origin of the source~\cite{eysenbach2008credibility}. This can impact the perceived credibility of messages and charts. Thus, we need to understand the effects of graphical dimensions that can provide the contextual information of charts and user characteristics on viewers' perceived credibility.

  When designing a visualization, visual designers have to make a decision on what information (e.g., title, historical context of the message) to include and make design decisions to represent that information (e.g., text with annotation to provide additional context~\cite{kong2018frames,stokes2022striking}, infographics). These design decisions can impact viewers' perception of credibility on charts. Prior studies have shown that embellished charts are likely to draw people's attention more and be more engaging, thus having viewers be more involved in understanding the message of the chart~\cite{wojdynski2015interactive,bateman2010useful}. But  we understand very little about how these graphical dimensions and design choices impact viewers' perceived chart credibility. Prior work has also shown that based on people's familiarity with the content impacts how people learn and read content~\cite{patel2018analysis,collins2002third,castles2018ending,national2000people}, which can impact people's data interpretation and perception chart credibility. Based on their knowledge level of the content, they might differentially perceive the credibility of charts. 

\section{Proposed Research \& Research Aims}

The core aim of this research is to understand and measure the effects of graphical dimensions and people's familiarity with the content on viewers' perceived chart credibility. 

\textbf{First, identify different graphical dimensions worth testing.} In my work, I plan to test text annotations, infographics, and traditional charts. These graphical dimensions were selected based on literature reviews. I also selected these test conditions because they are employed in realistic scenarios, such as in data journalism. In particular, I will test these on bar charts and line charts, as these are the basic types of charts that are commonly used.

\textbf{Second, measure the effects of graphical dimension and viewer's familiarity with the content on the viewer's perceived chart credibility.} I will design a study to measure viewers' perceived credibility of charts using credibility metrics(e.g., accuracy, fairness, trustworthiness)~\cite{gaziano1986measuring,meyer1988defining}. I will conduct crowd-sourced studies to assess them.

\textbf{Lastly, compose design guidelines for visual designers with experimental evidence.} We want to assess the effects of design choices on viewers' perceived chart credibility to propose chart design guidelines. 

\section{Planned Methodology}

To understand and assess the impact of graphical dimensions and users' familiarity with content on viewers' perceived chart credibility, I will conduct a crowd-sourced study (e.g., MTurk). The study will be a within-subjects study where participants will see all of the test conditions. I will use mixed methods to collect both quantitative and qualitative data. I will recruit two groups, subject domain experts and non-experts, based on their expertise level with the content. I will use a screener to determine their expertise level. Participants will be asked to read charts and be asked to self-report their perceived chart credibility using credibility metrics~\cite{gaziano1986measuring,meyer1988defining} (1 - very low, 7 - very high). Then I will collect why they reported those scales to understand how graphical dimensions or their expertise level impacted their subjective reportings.

\subsection{Hypotheses}

These are the set of hypotheses I'd like to test in my study.
\begin{itemize}[noitemsep,nolistsep]
    \item H1: Viewers will report the perceived chart credibility to be higher with richer annotations.
    \item H2: Viewers will report the perceived chart credibility to be lower for infographics than charts with annotations and traditional charts.
    \item H3: Viewers will report the infographics to be more engaging. 
\end{itemize}

\subsection{Study conditions}

I plan to test these conditions on bar charts and line charts as they are commonly used in various scenarios.
I will focus on two factors, 1) graphical dimensions and 2) the viewer's familiarity with the content. For graphical dimensions, I will test 1) traditional charts, 2) text annotations, and 3) infographics. For text annotation, I will use different semantic levels of text annotations~\cite{lundgard2021accessible}. These conditions were selected based on realistic scenarios, where these conditions are employed and considered when designing charts for data-driven journalism and storytelling tools. Also, for viewers' familiarity with the content, I will test with subject domain experts and non-experts. 

\section{Progress so far \& Next steps}

First, I identified graphical dimensions and user characteristics worth testing. 

Second, I defined perceived credibility and decided on potential tasks to measure viewers' perceived char credibility for my study. I also decided on potential data sets to use in my study.

Lastly, for the next step, I am refining the study design and plan to create or select visualizations to use in the crowdsourced studies. For basic charts and text annotation conditions, I will create them with d3.js or Vega lite. For infographics, I will use the already created versions of visualization.

\section{Challenges}
The following steps and challenges need to be addressed for this work to be successful. So, I look forward to getting feedback and input on the potential direction of the study and the potential experiment design would be useful.

First, Identify optimal data and tasks to use for the study. I am planning to use domains that are frequently used in data journalism, such as sports and politics. 
    
Second, identify tasks and how to evaluate perceived credibility for crowdsourced study.

\section{Expected Contributions}

My work will produce the following contributions to the research community:

\begin{itemize}[noitemsep,nolistsep]
    \item Study of effects of graphical dimensions and subject expertise viewers level on viewer's perceived chart credibility.
    \item Measurements of different graphical dimensions and user characteristics on viewer's perceived credibility on charts. 
    \item Chart design guidelines with experimental evidence that visual designers can refer to when making design decisions.
\end{itemize}



\bibliographystyle{abbrv-doi}

\bibliography{template}
\end{document}